\documentclass[prc,preprint,showpacs,nofootinbib]{revtex4}
\usepackage{graphicx}
\def\bge{\begin{equation}}
\def\ene{\end{equation}}
\def\bg{\begin{eqnarray}}
\def\en{\end{eqnarray}}

\begin{document}

\title{Magnetization of a neutron plasma with Skyrme and Gogny forces in the presence of a strong magnetic field.}

\author{M. \'Angeles P\'erez-Garc\'ia~\footnote{mperezga@usal.es}}

\affiliation{Department of Fundamental Physics and\\
Instituto Universitario de F\'isica Fundamental y Matem\'aticas \\
 University of Salamanca, Plaza de la Merced s/n, 37008, Salamanca, Spain}

\date{\today}

\begin{abstract}
Some thermodynamical magnitudes of interest in a pure neutron plasma are studied within the framework of the non-relativistic Brueckner-Hartree-Fock  approximation at finite density and temperature.  We use Skyrme and Gogny forces to describe such a neutron plasma and study the main differences that arise in these two effective parametrizations of the nuclear interaction when a strong magnetic field induces a permanent magnetization in the gas. The existence of a non-zero permanent spin polarization in a neutron plasma is explored in the density-temperature parameter space. We find that for moderate temperatures and in the low density range up to densities $\approx 0.5\rho_0$ both parametrizations predict that as density decreases an increasingly strong magnetization is allowed. In the range $0.5 \rho_0 \lesssim \rho \lesssim 3 \rho_0$ there is an approximately constant polarization that can be as big as $\approx 12\%$ for the maximum allowed interior magnetic field $B \approx 10^{18}$ G. For higher densities there is a dramatic difference in the polarization trend followed by Skyrme an Gogny forces. While the former predict  a ferromagnetic phase transition, the Gogny forces prevent it keeping the magnetization below $5\%$.
\vspace{1pc}
\end{abstract}

\maketitle

\section{Introduction}
\label{intro}
The presence of magnetic fields in neutron stars and their effect in the thermodynamics of plasmas in the inner regions of neutron stars and in proto-neutron stars has been a subject of study for long time, see ~\cite{woods} and references therein. Since the pioneering work of  Woltjer~\cite{woltjer} who predicted very intense magnetic fields in neutron stars as a result of flux conservation from the projenitor star until the recent experimental measurements of pulsar spin-down rates or X-ray spectral lines from protons inmersed in such intense magnetic fields~\cite{ibrahim} much work has been done. Experimental insight into the field has been obtained from meaurements ranging from radio pulsars where magnetic fields are believed to be in the range $B \approx 10^9-10^{12}$ G to the so called "magnetars" predicted to exist by Duncan and Thompson in 1992~\cite{duncan}  and first detected in 1998 ~\cite{kouveliotou} where $B_{magnetar}\approx 10^{14}-10^{15}$ G.  These field strengths strongly decay in a relatively rapid time of $10^3-10^4$  years. Two types of astronomical objects seem to be golden candidates for magnetars: soft gamma repeaters (SGRs) and anomalous X-ray pulsars (AXPs). Experimental observations confirm the idea that both are highly magnetized young neutron stars and can be linked to their original Supernova remnants~\cite{cline}.
Magnetars are believed to be around $10\%$ of the whole population of neutron stars so they are relatively frequent objects although only a handful of both have been observed~\cite{woods}. Other unconfirmed indications of strong magnetic fields include the possibility of an initial kick-off velocity due to an anisotropic neutrino-pair emission as suggested in~\cite{horli,arras}.
 
From a theoretical point of view, to study  how matter behaves in the presence of such strong magnetic fields some works have dealt with the possibility that nuclear plasmas may be partially magnetized~\cite{works_skyrme1,works_skyrme2,works_skyrme3,prakash1,prakash2,works_skyrme5}. It is widely known that Skyrme models have a characteristic pathology allowing a  ferromagnetic transition at high densities even in the absence of magnetic fields~\cite{works_skyrme4}. However, this transition is not confirmed when simulations are performed using realistic  interactions~\cite{fantoni}. In the outer shells of neutron stars one should focus in the low density limit of nuclear plasmas where clustered phases may develop~\cite{pasta1,pasta2,pasta3,pasta4}. For a magnetized system thus, we may have microscopic spin saturated spatial structures similar to magnetic domains. If present, they may largely  enhance the axial response in the neutrino opacities relevant to proto-neutron star early cooling. Since neutrino transport seems to be a crucial ingredient in the simulation of supernova core collapse it is important to size the effect of the existence of a strong magnetic field in the microphysics of these plasmas. 

The present work performs a comparative study of thermodynamical observables using Skyrme and Gogny effective nuclear interactions and focus on the possible existence of permanent imanation in the neutron plasma at finite temperature. The results show that, in the low density region, it is possible to have an increasingly stronger permanent polarization in the plasma at moderate temperatures as magnetic field strengths, higher than $B \approx 10^{16}$ G, exist. Below this strength the effects are almost negligible. At intermediate densities there is a plateau where both forces predict that the polarization remains approximately constant with maximum values around $12\%$. For higher densities, up to the limit where the nucleonic scenario holds, there is a dramatic difference in the behaviour predicted by these two forces. While Skyrme forces allow a ferromagnetic transition Gogny forces prevent it.
The structure of the paper is as follows. In section~\ref{model}, we describe the different nuclear  potentials used in the description of the nucleon-nucleon (NN) interaction, namely the Skyrme and Gogny forces. Later in this section we discuss the thermodynamical observables involved in the description of a nuclear plasma in the presence of a magnetic field. Results will be presented in section~\ref{results}, and summary and conclusions will be given in section~\ref{summary}.

\section{Thermodynamical potentials and nuclear interaction models}
\label{model}

We will start by describing the system we want to study. We consider a pure homogeneous neutron plasma which is composed of particles with spin projection on the  z-axis, $\sigma$. It can be either $\sigma=+1$ for spins aligned parallel to a uniform magnetic field that will be taken along the z-direction, ${\bf B}=B {\bf k}$, or $\sigma=-1$ for antiparallel spins. For a more complete description additional components in the hadronic and leptonic sector should be considered when beta equilibrium is imposed, however, for simplicity, we have restricted ourselves to the pure one component neutron plasma in this work.

The neutron plasma density, $\rho$, is composed of up($+$) and down($-$) spin aligned particles  
\begin{equation}
\rho=\rho_{+}+\rho_{-}
\label{dens}
\end{equation}
where 
\begin{equation}
\rho_{\sigma}=\int \frac{d^3 k}{(2 \pi)^3} f_{\sigma}(k)
\label{densig}
\end{equation}
and the $f_{\sigma}(k)$ is the Fermi distribution function for the fermion component with spin projection $\sigma$ at temperature $T=1/\beta$ ($k_B=1$)

\begin{equation}
f_{\sigma}(k)=\frac{1}{1+e^{\beta(\epsilon_{\sigma}(k)-\mu)}}
\label{nsigma}
\end{equation}
The single particle energy $\epsilon_{\sigma}(k)$ will be especified for each interaction model that will be used in this work, see below.
The magnetization in the neutron plasma is defined as

\begin{equation}
M=\mu_n \Delta \rho
\end{equation}

where $\mu_n=-1.9130427(5)\mu_N$ is the neutron magnetic moment in units of the nuclear magneton~\cite{pdb} and $\Delta$ is the spin excess or polarization of the system
\begin{equation}
\Delta =\frac{\rho_{+}-\rho_{-}}{\rho}
\label{delta}
\end{equation}

\subsection{Skyrme force}
\label{Skyforce}
We have considered, in first place, the phenomenological Skyrme interaction that appears in the literature under a rather general form~\cite{vautherin} 
\begin{eqnarray}
V^{Skyrme}_{NN}({\bf r}_1,{\bf r}_2)&=& t_0 \left(1+x_0 P^{\sigma} \right) \delta({\bf r})  + \frac{1}{2} t_1 \left(1+x_1 P^{\sigma} \right)
\left[ {\bf k'}^2 \delta({\bf r}) +  \delta({\bf r}) {\bf k}^2 \right] 
\nonumber \\
&& +  t_2 \left(1+x_2 P^{\sigma} \right) {\bf k'} \cdot \delta({\bf r}) {\bf k}
+ \frac{1}{6} t_3 \left(1+x_3 P^{\sigma} \right) \rho^{\alpha} ({\bf R})\delta({\bf r}) 
\label{skyrme}
\end{eqnarray}
where ${\bf r}={\bf r}_1-{\bf r}_2$ and ${\bf R}=({\bf r}_1+{\bf r}_2)/2$, 
${\bf k}=({\bf \nabla}_1-{\bf \nabla}_2)/2i$ the relative momentum acting on the right and ${\bf k'}$ its conjugate acting on the left. $P^{\sigma}$ is the spin exchange operator. Note that we have omitted terms not relevant for homogeneous systems.
As widely known, this potential allows for a good reproduction of finite nuclei and their excited states and also bulk matter relevant to neutron stars~\cite{chabanat, chabanat2, skyrme2,skyrmeeos}.

In order to study the neutron plasma we must consider the partition function, $\mathcal{Z}$,  of the available configuration space, $\Omega$. The one that must be used for our system with hamiltonian $\mathcal{H}$ is that of the grand canonical ensamble under the influence of an external magnetic field. 

\begin{equation}
\mathcal{Z}({\Omega})=e^{-\beta F_M(\Omega)}=Tr \{e^{-\beta \mathcal{H}(\Omega)}\}
\end{equation}

The thermodynamical potential relevant for the description of a plasma at temperature T and external magnetic field $H$ is the Helmholtz free energy, $F_M$, defined as~\cite{callen}  
\begin{equation}
F_M=E-TS-MH
\label{Fm}
\end{equation}

Note that previously we have used $B$ to designate the magnetic field strength, however one should have in mind that apart from the external currents providing the external field $H$, the magnetization may contribute to the internal magnetic field~\cite{landau}
\begin{equation}
B=H+4 \pi M
\label{Hfield}
\end{equation}
A self-consistent calculation would require considering the contribution of the magnetization to the total magnetic field $B$, however we expect $M$ to be much less than the external field, so we will consider that the ratio $\mid H/B \mid$ will always be close to unity~\cite{prakash1}. In what follows we will keep the notation using $B$ to designate the total magnetic field strength. In addition we will use capital letters for any extensive magnitude, A, and, for its associated intensive magnitude (per particle), the same letter in lower-case, a=A/N.

Using the non-relativistic Brueckner-Hartree-Fock approximation, the single particle energy for a neutron with momentum $k$ and spin projection $\sigma$ in presence of a magnetic field $B$ can be written as
\begin{equation}
\epsilon_{\sigma}(k)=\frac{\hbar^2 k^2}{2 m} +U_{\sigma}(k)+\mu_n \sigma B
\label{singlepart2}
\end{equation}

where $U_{\sigma}(k)$ is the single particle potential~\cite{works_skyrme5}. For the Skyrme interaction it has a quadratic k-dependence that can be included into an effective mass plus a k independent term, $\overline{U}_{\sigma}$, as
\begin{equation}
\epsilon_{\sigma}=\frac{\hbar^2 k^2}{2 m^*_{\sigma}} +\overline{U}_{\sigma}+\mu_n \sigma B
\label{singlepart}
\end{equation}
where the effective mass is given explicitly by
\begin{eqnarray}
\frac{ m^*_{\sigma}}{m} &=& \frac{1}{1+\frac{2m}{(\hbar c)^2}\left[
\frac{1}{4} (t_1 (1-x_1) + t_2 (1+x_2)) \rho_{-\sigma} +2 t_2 (1+x_2)\rho_{\sigma}\right]}
\label{meff_skyrme}
\end{eqnarray}
that in this approach is independent of temperature. It is useful at this point, in order to calculate the thermodynamical quantities of interest in the magnetized plasma, to define an effective neutron chemical potential into the following form
\begin{equation}
\nu_{ef\,\sigma}=\mu-\overline{U}_{\sigma}-\mu_n \sigma B
\label{nuef_skyrme}
\end{equation}

Then, the Fermi distribution reads like that of a free Fermi gas with effective mass $m^*_{\sigma}$ and chemical potential $\nu_{ef\,\sigma}$.

\begin{equation}
f_{\sigma}(k)=\frac{1}{1+e^{\beta(\frac{\hbar^2 k^2}{2 m^*_{\sigma} }-\nu_{ef\,\sigma})}}
\label{nsigma1}
\end{equation}

However, the single particle potential will be in general a function of $k$, ${U_{\sigma}(k)}$, as {\it i.e.} when using the Gogny force. See later in this section.

In the plasma the total energy per particle, for a given interaction model, can be obtained from averaging the kinetic, potential and field contributions

\begin{equation}
e=\frac{1}{\rho} \sum_{\sigma}\int \frac{d^3 k}{(2 \pi)^3} \frac{\hbar^2 k^2}{2 m} f_{\sigma}(k)+\frac{1}{\rho} \sum_{\sigma,\sigma'} \int \frac{d^3 k d^3 k'}{(2 \pi)^6} \langle V_{NN}\rangle_A f_{\sigma}(k) f_{\sigma'}(k')+\frac{B^2}{8 \pi \rho}
\label{e}
\end{equation}

For the Skyrme interaction it can be casted into the following form:
\begin{eqnarray}
{e}^{Skyrme}&=&\frac{1}{\rho} \frac{\hbar^2}{2 m^*_{+}}\tau_+ + \frac{1}{\rho} \frac{\hbar^2}{2 m^*_{-}}\tau_-+ \nonumber \\
&& \frac{1}{\rho} 
\left[ t_0 (1-x_0) + \frac{1}{6} t_3 (1-x_3) \rho^{\alpha} \right] \rho_+ \rho_- +\frac{B^2}{8 \pi \rho}
\label{e_skyrme}
\end{eqnarray}
where $\tau_{\sigma}$ can be expressed as
\begin{equation}
\tau_{\sigma}=\int \frac{d^3 k}{(2 \pi)^3} k^2 f_{\sigma}(k)
\label{tausigma}
\end{equation}
We have considered in this work two Skyrme parametrizations given by the Lyon group, namely we have chosen the SLy4 and SLy7 parametrizations~\cite{chabanat,chabanat2} as they are the most commonly used in the literature. These provide good values for binding of nuclei and neutron matter equation of state (EOS) giving values of maximum neutron star masses around $1.5 M_{\odot}$. In Table~\ref{tab:param} we summarize the values of some observables: saturation density, $\rho_0$, binding energy for symmetric nuclear matter, $a_v$,  symmetry energy, $a_s$,  and incompressibility modulus, $K_{\infty}$, for the effective interaction models used in this work.
\begin{table}[htbp]
\begin{center}
\caption{Values of some observables with the Skyrme and Gogny forces considered in this work~\cite{chabanat, chabanat2}~\cite{d1s, d1p}.}
\label{tab:param}
\begin{tabular}{l|llll} \hline
Model &$\rho_0 (fm^{-3})$   &$K_{\infty}$ (MeV) & $a_v$ (MeV) & $a_s$ (MeV)\\ \hline
SLy4 &0.160  &230.9 &-15.97 &32.00 \\
SLy7 &0.158  &229.7 &-15.89  &31.99 \\
D1S &0.1625  &203   &-16.01 &31.13\\
D1P &0.1737  &266  &-16.19 &34.09
\end{tabular}
\end{center}
\end{table}

\subsection{Gogny force}
\label{gogny}

The Gogny force has been extensively used in the literature not only for describing finite nuclei but for the EOS of pure neutron matter~\cite{works_gogny3,works_gogny4,works_gogny1,works_gogny2}.
The interaction potential is given by the contribution of finite range terms and zero-range terms:
\begin{eqnarray}
V^{Gogny}_{NN}({\bf r}_1,{\bf r}_2)&=& \sum_{i=1}^{2} \{ [W_i + B_i P^{\sigma} -
H_i P^{\tau} - M_i P^{\sigma} P^{\tau} ] e^{ -|{\bf r}_1-{\bf r}_2|^2 / \mu_i^2 }+ \nonumber \\ 
&& t_{3i}(1+x_3 P^{\sigma} ) \rho^{\alpha_i}\delta({\bf r})  \}
\label{v_gogny}
\end{eqnarray}
where $P^{\sigma}$ and $P^{\tau}$ are the spin and isospin exchange operators. The values of some observables for the two parametrizations considered in this work for the Gogny force appear in Table I. The D1S parametrization was originally introduced to describe the pairing properties and surface effects in finite nuclei~\cite{d1s} while the D1P  aims at reproducing the EOS of neutron matter obtained using realistic interactions~\cite{d1p}. 

In this model the expression for the single particle potential felt by a neutron of momentum k and spin projection $\sigma$ is given by
\begin{eqnarray}
U^{Gogny}_{\sigma}(k)&=& \sum_{i=1}^{2} \{ t_{3i} (1-x_0) \rho^{\alpha_i} \rho_{-\sigma}+ \alpha_i t_{3i} \rho^{\alpha_i-1} \} + 
\pi^{3/2} \sum_{i=1}^{2} \mu^3_i \{[ W_i-H_i ] \rho + \nonumber \\ 
&& [B_i-M_i]  \rho_{\sigma}-[ W_i+B_i-H_i-M_i ]I^{\sigma}_i(k,T)-[ B_i-M_i ]I^{-\sigma}_i(k,T )\}
\label{u_gogny}
\end{eqnarray}

where the value of the $I^{\sigma}_i$ integral is
\begin{equation}
I^{\sigma}_i (k,T)=\int \frac{d^3k'}{(2 \pi)^3} f_{\sigma}(k') 
e^{-\frac{\mu^2_i}{4} |{\bf k}-{\bf k'}|^2}
\label{I_i}
\end{equation}
The integral $I^{- \sigma}_i$ in Eq. (\ref{u_gogny}) must be calculated using the oppositely polarized particle distribution $f_{-\sigma}$. For the Gogny force we must note that the single particle energy entering the distribution function $f_{\sigma}$  is given by 
\begin{equation}
\epsilon_{\sigma}(k)=\frac{\hbar^2 k^2}{2 m}+U^{Gogny}_{\sigma}(k)+\mu_n \sigma B
\label{ener_gogny}
\end{equation}

Recent works~\cite{magnetized} have shown that nucleon effective masses in plasmas described with Gogny forces have a weak temperature dependence thus, as an approximation, the most physically relevant interactions may be considered to take place around the Fermi surfaces with momentum $k_{F\,{\sigma}}$. Then, we will consider here a quadratic approximation in k-space to the full single particle potential since it proves to be a good approximation at least for $k$ up to values around $k\approx k_{F\,{\sigma}}$ as shown in ref.~\cite{polls1}.
In this way the quadratic momentum approximation for the single particle potential can be written as 
\begin{equation}
U^{Gogny}_{\sigma}(k)\approx U^{Gogny}_{\sigma}(k=0)+\left (\frac{1}{2 k} \frac{d U^{Gogny}_{\sigma}((k)}{dk}\right )_{k=k_{F\,\sigma}} k^2
\label{U2}
\end{equation}

and an effective mass can be defined as
\begin{equation}
\frac{m^*_{\sigma}}{m}=\left(1+  \frac{m } {\hbar^2 k'} \frac{dU^{Gogny}_{\sigma}(k')}{dk'} \right)^{-1}_{k'=k}
\label{effm}
\end{equation}

To obtain the thermodynamical variables of interest once temperature, density, magnetic field strength and imanation are fixed,  we must solve selfconsistently the set of equations given by
\begin{equation}
\frac{{m^*}_{\sigma}}{m}=\left (1+ \frac{m}{\hbar^2 k} \frac{dU^{Gogny}_{\sigma}(\rho, k', T, B)}{dk'}\right )^{-1}_{k'=k_{F_{\sigma}}}
\label{eq1}
\end{equation}
\begin{equation}
\rho=\rho_{+}+\rho_{-}
\label{eq2}
\end{equation}
\begin{equation}
\Delta \rho=\rho_{+}-\rho_{-}
\label{eq3}
\end{equation}

Above, Eq.~(\ref{eq1}) represents the mean field felt by the up (down) particles in the plasma while Eq.~(\ref{eq2}) must be solved in order to fix neutron number conservation and Eq.~(\ref{eq3}) relates to the imanation constraints in the plasma. When analyzing the case of the Skyrme force the expression of the effective mass is analytical and Eq.~(\ref{eq1}) must be substituted by Eq.~(\ref{meff_skyrme}).

Once this set of equations is solved, we must evaluate other thermodynamical quantities on the 
hyper-surface $\Delta(\rho, T, B)$. The energy per particle is given by 
\begin{eqnarray}
e^{Gogny}&=&  \frac{1}{\rho} \frac{\hbar^2}{2 m}(\tau_+ +\tau_-)+ \nonumber \\
&&\sum_{i=1}^{2} \frac{1}{\rho} t_{0i}(1-x_0) \rho^{\alpha_i-1} \rho_+ \rho_- +
\pi^{3/2} \mu_i^{3}\{[W_i-H_i]\frac{\rho}{2}+\frac{\rho_+^2+\rho_-^2}{2 \rho} [B_i-M_i]- \nonumber \\
&& [W_i+B_i-H_i-M_i] \frac{1}{2 \rho} [G_i^{(+,+)}(\rho,T)+G_i^{(-,-)}(\rho,T)]-
\nonumber \\
&& [B_i-M_i] \frac{1}{2 \rho} [ G_i^{(+,-)}(\rho,T)+G_i^{(-,+)}(\rho,T) ]\}+ \frac{B^2}{8 \pi \rho}
\label{e_gogny}
\end{eqnarray}
where the value of the $G_i$ integral is
\begin{equation}
G_i^{(\sigma,\sigma')}(\rho,T)=\int \frac{d^3k}{(2 \pi)^3} f_{\sigma}(k,T) I_i^{\sigma'}(k,T)
\label{g_i}
\end{equation}

The entropy per particle can be written then as 

\begin{equation}
s= \frac{-1}{\rho} \sum_{\sigma=-,+} \int \frac{d^3k}{(2\pi)^3} f_{\sigma}(k)ln(f_{\sigma}(k))-(1-f_{\sigma}(k))ln(1-f_{\sigma}(k))
\label{entropy}
\end{equation}

To extract the energetically favoured spin polarization in this ensamble the free energy per particle, $f_M$, has to reach a minimum, $\Delta_{min}$, on the hypersurface  at a given density, temperature and magnetic field strength. 
\begin{equation}
\left( \frac{\partial f_M}{\partial \Delta}\right)_{\rho, T, B}=0
\label{deriv_f}
\end{equation}

\section{Results}
\label{results}

In this section we present the results. We have considered magnetic fields up to a maximum value  according to the scalar virial theorem. For a neutron star of radius $R$ the maximum allowed stored magnetic energy is given by the relation $\frac{4 \pi R^3}{3}\frac{B_{max}^2}{8 \pi} \approx \frac{G M^2}{R}$ then, for a  typical neutron star with $M=1.5 M_{\odot}$ and $R=10^{-5} R_{\odot}$ this yields an interior maximum value of the magnetic field $B_{max} \approx 10^{18}$ G. Recent observations report surface values up to $B \approx 10^{15}$ G as deduced from hydrogen spectral lines~\cite{ibrahim} but even higher magnetic field strengths have been speculated~\cite{chakra1}. 

\begin{figure}[hbtp]
\begin{center}
\includegraphics [angle=-90,scale=.75] {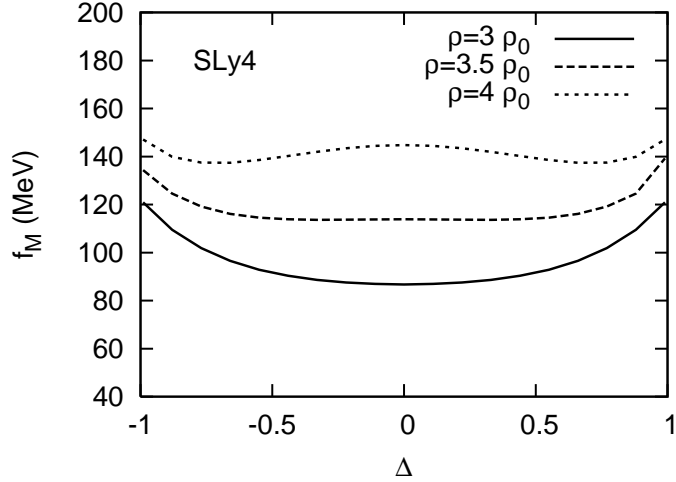}
\caption{Free energy per particle in absence of magnetic field for different values of density and $T=2$ MeV with the Skyrme SLy4 model as a function of the polarization.} 
\label{Fig14}
\end{center}
\end{figure}

Skyrme models show a pathology, a ferromagnetic transition in the range $[1-4]\rho_0$, that is due to the appearance of a spontaneous polarization in the system even if no magnetic fields are present. 
To illustrathe this we show in Fig.~\ref{Fig14} the development of a non-zero value of polarization $\Delta$ where the free energy per particle achieves a minimum until a ferromagnetic transition occurs around $\approx 4\rho_0$ for the SLy4 model at a temperature $T=2$ MeV and vanishing magnetic field strength.
%%%%%%%%%%%%%%%%%%%%%%%%%%%%%%%%%%%%%%%%%%%%%%%%%%%%%%%%%%%%%%%%%

\begin{figure}[hbtp]
\begin{center}
\includegraphics [angle=-90,scale=.75] {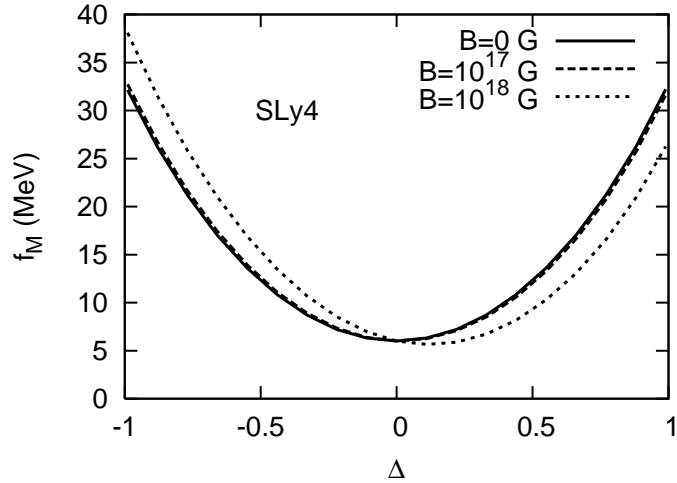}
\caption{Free energy per particle at a density $\rho=0.5 \rho_0$ for different values of magnetic field at $T=10$ MeV with the Skyrme SLy4 model as a function of the polarization.}
\label{Fig15}
\end{center}
\end{figure}

In Fig.~\ref{Fig15} we show the free energy per particle for a plasma at $\rho=0.5\rho_0$ for a typical temperature in a proto-neutron star, $T=10$ MeV, and three different values of the magnetic field strength as calculated with the SLy4 parametrization. We can see that for fields below $B\approx 10^{17}$ G the effects introduced by the presence of a magnetic field are almost negligible. Then, when including the magnetic field the polarization symmetry is broken and no longer appear two minima.
%%%%%%%%%%%%%%%%%%%%%%%%%%%%%%%%%%%%%%%%%%%%%%%%%%%%%%%%%%%%%%%%%
%%%%%%%%%%%%%%%%%%%%%%%%%%%%%%%%%%%%%%%%%%%%%%%%%%%%%%%%%%%%%%%%%

\begin{figure}[hbtp]
\begin{center}
\includegraphics [angle=-90,scale=.75] {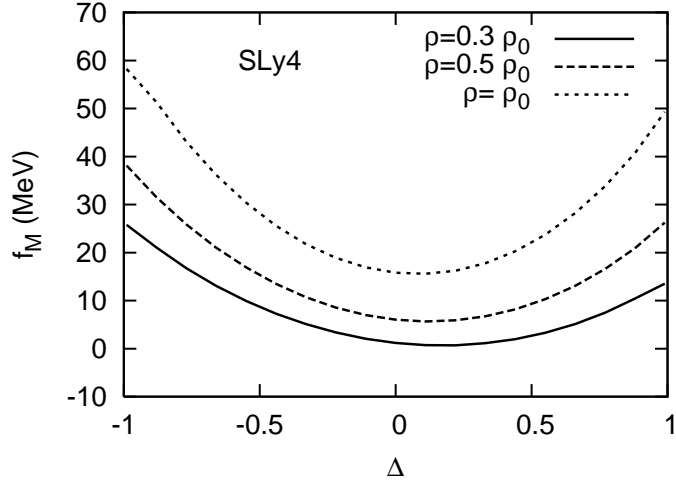}
\caption{Free energy per particle for different values of the density at $T=10$ MeV with the Skyrme SLy4 model and $B=10^{18}$ G as a function of the polarization.} 
\label{Fig16}
\end{center}
\end{figure}

In the low density region the free energy is an increasing function of the density but the minimun of the polarization shifts to higher values as can be seen in Fig.~\ref{Fig16} for $T=10$ MeV and $B=10^{18}$ G using the SLy4 model. At densities below $0.5 \rho_0$ a wide range of models show the development of a frustrated state of matter that is known with the name of {\it pasta}~\cite{pasta1,pasta2,pasta3,pasta4}. Different simulations performed without the presence of a magnetic field show a variety of spatial shapes that form in neutron rich matter. It is important to explore the thermodynamical conditions for magnetized pasta where the presence of magnetic domains could largely impact neutrino opacities in the proto-neutron star early cooling stage~\cite{future}.
%%%%%%%%%%%%%%%%%%%%%%%%%%%%%%%%%%%%%%%%%%%%%%%%%%%%%%%%%%%%%%%%%
%%%%%%%%%%%%%%%%%%%%%%%%%%%%%%%%%%%%%%%%%%%%%%%%%%%%%%%%%%%%%%%%%

\begin{figure}[hbtp]
\begin{center}
\includegraphics [angle=-90,scale=.75] {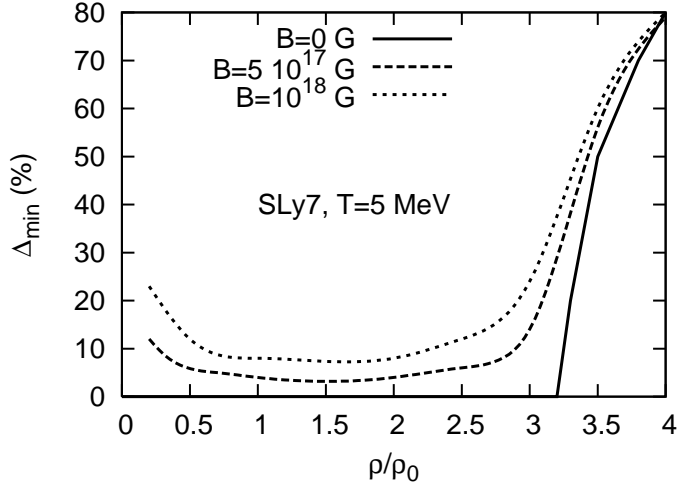}
\caption{Permanent polarization for different values of  the magnetic field strength at $T=5$ MeV with the Skyrme SLy7 model as a function of density.}
\label{Fig18}
\end{center}
\end{figure}

In Fig.~\ref{Fig18}  we show the polarization, in percentage ($\%$), that minimizes the free energy at $T=5$ MeV resulting from a calculation with the SLy7 model as a function of density for $\rho \le 4 \rho_0$. For higher densities a nucleon-quark transition is allowed by phenomenological models, see for example ~\cite{quark}, so we will restrict our non-relativistic analysis to this range. We can see that there are three differentiated polarization regions. At densities increasingly smaller below $\approx 0.5\rho_0$, the polarization rises  with increasing magnetic field strength. It is worth mentioning that we have assumed that even at these low densities the gas picture still holds. However, the formation of fully spin saturated microscopic nuclear regions or "domains" should be explicitly explored in the future~\cite{preparation}.
For densities in the range $0.5\rho_0 \lesssim \rho \lesssim 3\rho_0$ a stable permanent imanation of the neutron plasma is allowed, if a magnetic field is present, giving a maximun polarization around $12 \%$ for the maximun $B$ considered in this work. For densities  larger than $3\rho_0$, again, there is a very rapid increase of the spin polarization that achieves full saturation and therefore a ferromagnetic transition occurs at around $4\rho_0$ for this particular parametrization.

%%%%%%%%%%%%%%%%%%%%%%%%%%%%%%%%%%%%%%%%%%%%%%%%%%%%%%%%%%%%%%%%%
%%%%%%%%%%%%%%%%%%%%%%%%%%%%%%%%%%%%%%%%%%%%%%%%%%%%%%%%%%%%%%%%%

\begin{figure}[hbtp]
\begin{center}
\includegraphics [angle=-90,scale=.75] {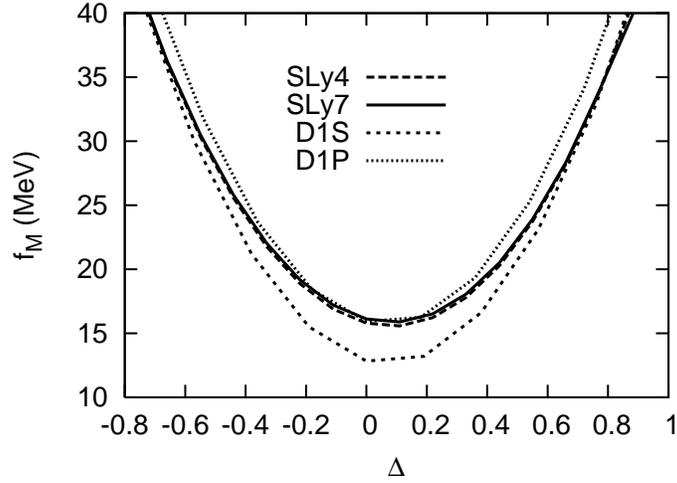}
\caption{Free energy per particle at a density $\rho=\rho_0$, $T=10$ MeV and $B=10^{18}$ G with different Skyrme and Gogny models as a function of the polarization.} 
\label{Fig17}
\end{center}
\end{figure}

\begin{figure}[hbtp]
\begin{center}
\includegraphics [angle=-90,scale=.75] {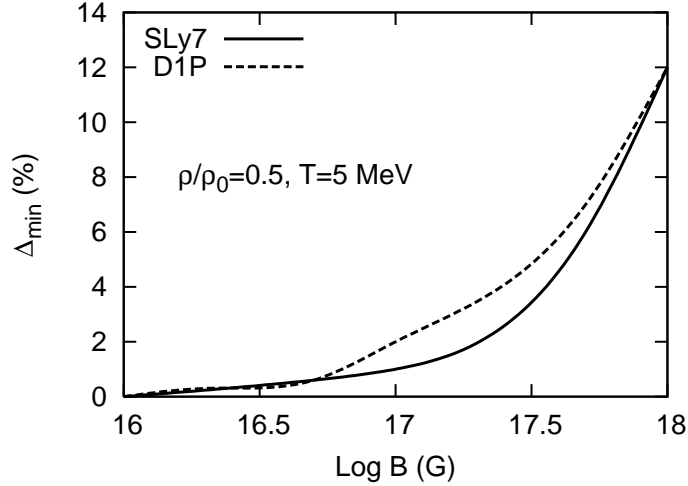}
\caption{Polarization at a density $\rho=0.5 \rho_0$ and  $T=5$ MeV with the Skyrme SLy7 and Gogny D1P model as a function of the logarithm of the magnetic field strength.} 
\label{Fig20}
\end{center}
\end{figure}

The robustness of the findings beforementioned can be seen on Fig.~\ref{Fig17} where a comparison of Skyrme (SLy4 and Sly7) and Gogny models (D1S and D1P) is shown for the case with $\rho=\rho_0$ at $T=10$ MeV and $B=10^{18}$ G. We can see that all of them predict the same value of the minimun around $\Delta_{min} \approx 12\%$. The detail of the robustness at low densities with respect to the Skyrme (SLy7) and Gogny (D1P) parametrizations is shown in Fig.~\ref{Fig20} for a density $\rho=0.5 \rho_0$ at a $T=5$ MeV where the permanent polarization is plot versus the logarithm of $B$. Below $B=10^{16}$ G the effect is negligible but it largely increases up to a value of $\approx 12 \%$ for the $B_{max}$. We can see that the increase of magnetization for sufficiently high magnetic field grows more rapidly than in the very low density and moderate magnetic field strengths regime where Curie's linear law $M \approx B/T$ is expected to hold.

%%%%%%%%%%%%%%%%%%%%%%%%%%%%%%%%%%%%%%%%%%%%%%%%%%%%%%%%%%%%%%%%%
%%%%%%%%%%%%%%%%%%%%%%%%%%%%%%%%%%%%%%%%%%%%%%%%%%%%%%%%%%%%%%%%%

%%%%%%%%%%%%%%%%%%%%%%%%%%%%%%%%%%%%%%%%%%%%%%%%%%%%%%%%%%%%%%%%%

%%%%%%%%%%%%%%%%%%%%%%%%%%%%%%%%%%%%%%%%%%%%%%%%%%%%%%%%%%%%%%%%%
\begin{figure}[hbtp]
\begin{center}
\includegraphics [angle=-90,scale=.75] {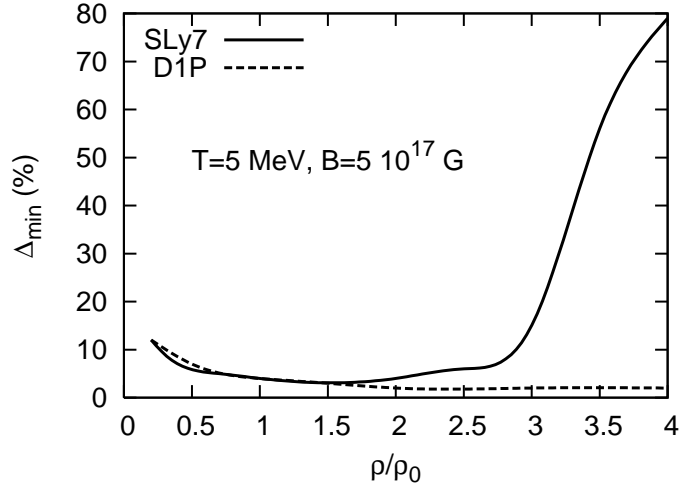}
\caption{Permanent polarization at $T=5$ MeV and $B=5\, 10^{17}$ G with the Skyrme SLy7 and Gogny D1P model as a function of the density.}
\label{Fig21}
\end{center}
\end{figure}
%%%%%%%%%%%%%%%%%%%%%%%%%%%%%%%%%%%%%%%%%%%%%%%%%%%%%%%%%%%%%%%%%

In Fig.~\ref{Fig21} we show the plasma polarization at $T=5$ MeV and $B=5\, 10^{17}$ G with the Skyrme SLy7 (solid line) and Gogny D1P (dashed line) models as a function of the density. It is clearly observed the general feature of enhanced polarization at low densities and a plateau at intermediate densities. However at high densities there is a dramatic difference since in the Skyrme model there is a ferromagnetic transition at $\approx 4\rho_0$ while it is forbidden with the D1P model in agreement with realistic simulations~\cite{fantoni}. In the case of the Gogny force there is a plateau with permanent residual polarization of about $2\%$. These results agree with those previously obtained with calculations performed in a relativistic  scenario in the context of a mean field approximation where they estimate a maximum magnetization smaller than $4\%$~\cite{prakash1}.

\begin{figure}[hbtp]
\begin{center}
\includegraphics [angle=-90,scale=.75] {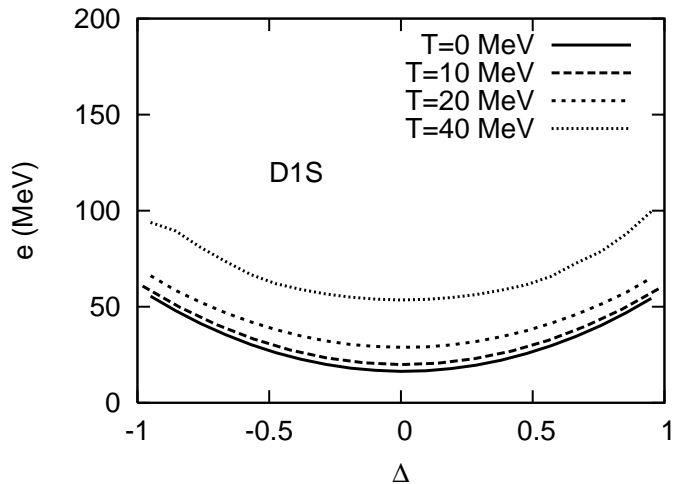}
\caption{Energy per particle at a density $\rho=\rho_0$ for different values of temperature and $B=10^{17}$ G with the Gogny D1S model as a function of the polarization. } 
\label{Fig12}
\end{center}
\end{figure}

\begin{figure}[hbtp]
\begin{center}
\includegraphics [angle=-90,scale=.75] {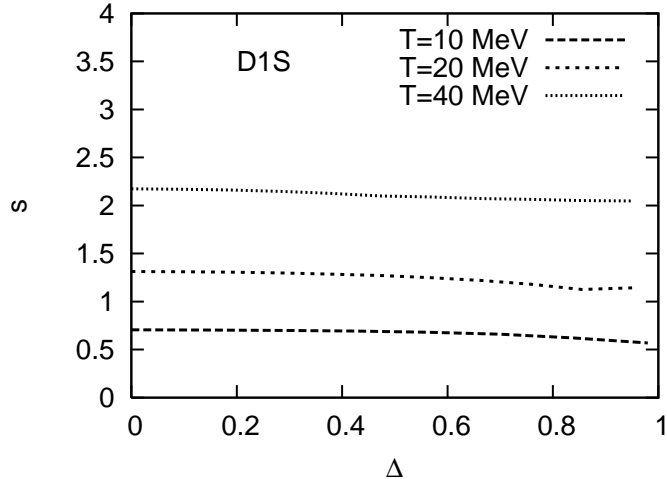}
\caption{Entropy per particle, same thermodynamical conditions as  in Fig.~\ref{Fig12}.} 
\label{Fig13}
\end{center}
\end{figure}

\begin{figure}[hbtp]
\begin{center}
\includegraphics [angle=-90,scale=.75] {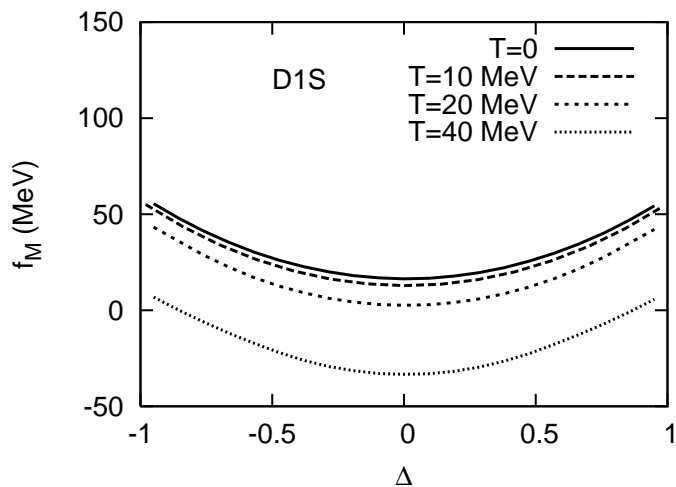}
\caption{Free energy per particle, same conditions as  in Fig.~\ref{Fig12}.}
\label{Fig11}
\end{center}
\end{figure}

The effects of temperature on other thermodynamical quantities of interest can be seen in Fig.~\ref{Fig12}, Fig.~\ref{Fig13} and Fig.~\ref{Fig11} where the energy, entropy and free energy per particle, respectively, are shown for a density $\rho=\rho_0$ at $B=10^{17}$ G at several temperatures for the D1S Gogny force. The typical range of temperatures relevant to proto-neutron stars goes up to values around $T=40$ MeV. We can see that there is a moderate dependence of the energy per particle, $e$, with temperature. In the case of the entropy we can see the familiar decreasing behaviour~\cite{ngo} with increasing order of the system, having a maximum around vanishing polarization. At high temperatures where correlations are greatly diminished one should recover the Sackur-Tretode formula limit $s\approx 2.5$~\cite{ngo}.

%%%%%%%%%%%%%%%%%%%%%%%%%%%%%%%%%%%%%%%%%%%%%%%%%%%%%%%%%%%%%%%%%
%%%%%%%%%%%%%%%%%%%%%%%%%%%%%%%%%%%%%%%%%%%%%%%%%%%%%%%%%%%%%%%%%
%%%%%%%%%%%%%%%%%%%%%%%%%%%%%%%%%%%%%%%%%%%%%%%%%%%%%%%%%%%%%%%%%
%%%%%%%%%%%%%%%%%%%%%%%%%%%%%%%%%%%%%%%%%%%%%%%%%%%%%%%%%%%%%%%%%

%%%%%%%%%%%%%%%%%%%%%%%%%%%%%%%%%%%%%%%%%%%%%%%%%%%%%%%%%%%%%%%%%
%%%%%%%%%%%%%%%%%%%%%%%%%%%%%%%%%%%%%%%%%%%%%%%%%%%%%%%%%%%%%%%%%

\begin{figure}[hbtp]
\begin{center}
\includegraphics [angle=-90,scale=.75] {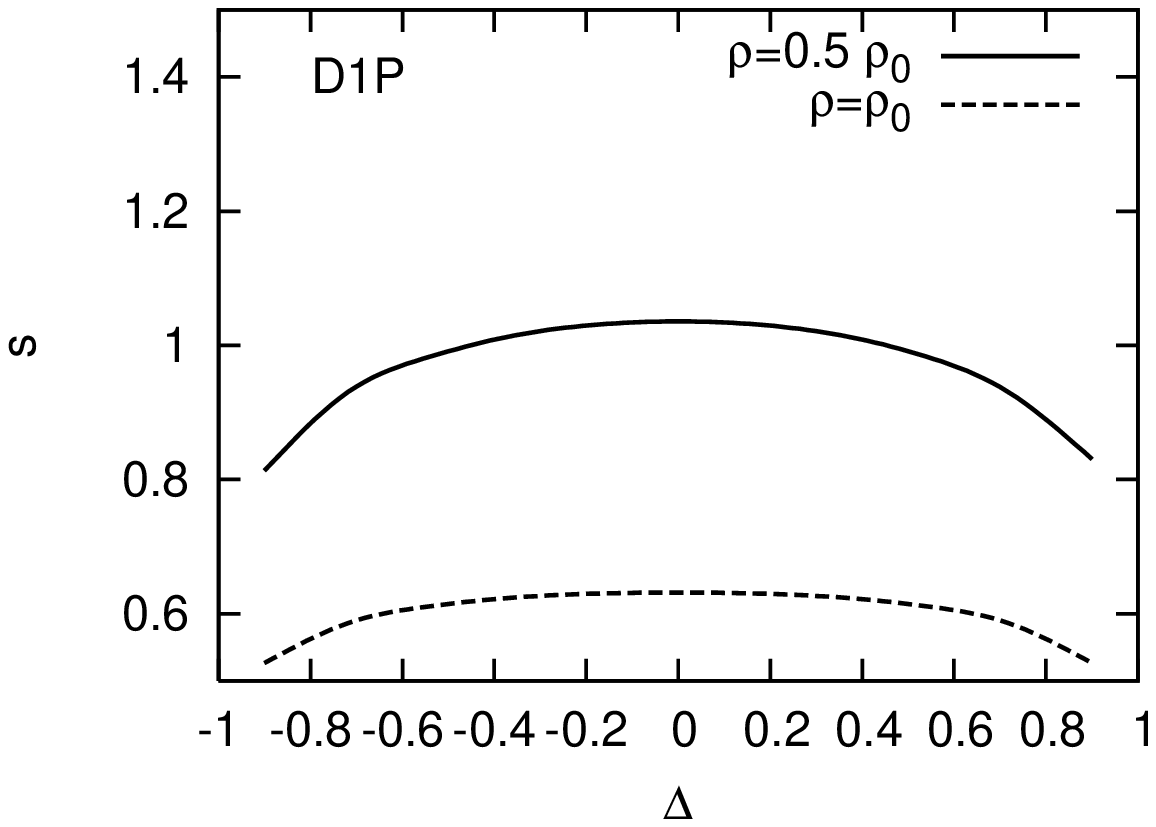}
\caption{Entropy per particle for $\rho=0.5 \rho_0$ (solid line) and $\rho=\rho_0$ (dashed line) at $B=10^{18}$ G and $T=10$ MeV with the Gogny D1P model.} 
\label{Fig4}
\end{center}
\end{figure}

In Fig.~\ref{Fig4} we show the entropy per particle as a function of the polarization for two different densities. The solid line corresponds to a density $\rho=0.5\rho_0$ and the dashed line to $\rho=\rho_0$  at $T=10$ MeV with the Gogny D1P model. For this plot we have taken the upper limiting case $B=10^{18}$ G. We can see that as density decreases in-medium effects become less important and allow the system more available configurations thus entropy increases. 

%%%%%%%%%%%%%%%%%%%%%%%%%%%%%%%%%%%%%%%%%%%%%%%%%%%%%%%%%%%%%%%%%
\begin{figure}[hbt]
\begin{center}
\includegraphics [angle=-90,scale=0.75] {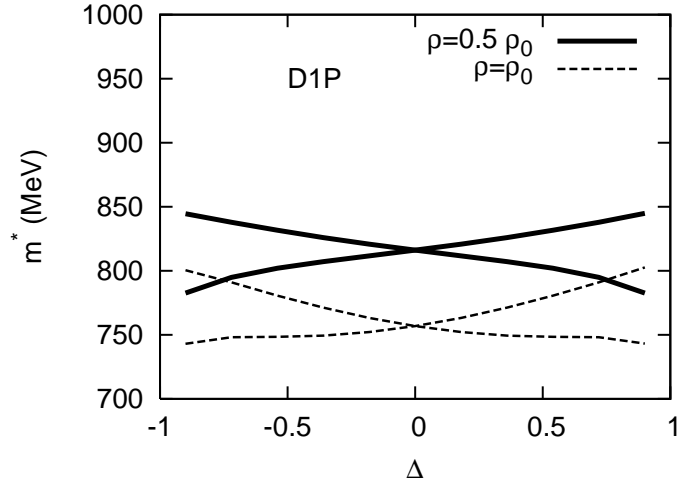}
\caption{Neutron effective mass for $\rho=0.5 \rho_0$ (solid line) and $\rho=\rho_0$ (dashed line) at $T=10$ MeV with the Gogny D1P model. The ascendent (descedent) curves are for up (down) polarized particles.}
\label{Fig3}
\end{center}
\end{figure}

\begin{figure}[hbtp]
\begin{center}
\includegraphics [angle=-90,scale=.75] {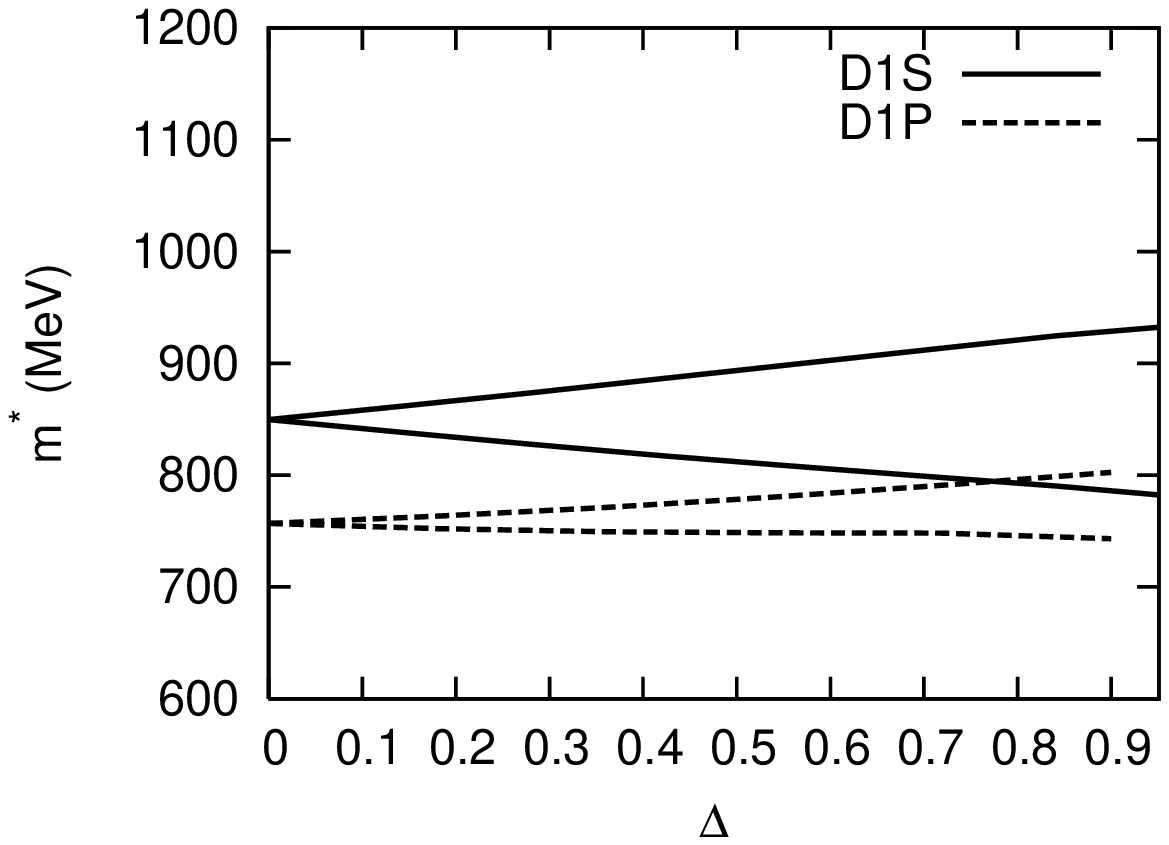}
\caption{Effective neutron mass at a density $\rho=\rho_0$ with the D1S model (solid line) and with the D1P model (dashed line) at $T=10$ MeV as a function of the polarization. The ascendent(descent) curves are for up (down) polarized particles.} 
\label{Fig9}
\end{center}
\end{figure}

%%%%%%%%%%%%%%%%%%%%%%%%%%%%%%%%%%%%%%%%%%%%%%%%%%%%%%%%%%%%%%%%%
The effective mass for up and down polarized neutrons is shown in Fig.~\ref{Fig3} for densities $\rho=0.5 \rho_0$ (solid line) and $\rho=\rho_0$ (dashed line) at $T=10$ MeV with the Gogny D1P model. As polarization increases, the ascendent (descendent) curves for each density correspond to up (down) polarized neutrons. As shown, in-medium effects decrease the effective neutron mass as density increases. In Fig.~\ref{Fig9} we show the effective neutron masses at a density $\rho=\rho_0$ with the D1S model (solid line) and with the D1P model (dashed line) at $T=10$ MeV as a function of the polarization. We can see that the effects due to the parametrization or in-medium correlations are much more important than those related to the polarization since for the maximum B considered in this work induces a $\Delta_{min} \approx 12\%$ that relates to a mass shift of an amount of $\delta m^* \approx 5 \, MeV$ with respec to the unpolarized case, and $\delta m^* \approx 10 \, MeV$ for the $\rho=0.5 \rho_0$ case.

%%%%%%%%%%%%%%%%%%%%%%%%%%%%%%%%%%%%%%%%%%%%%%%%%%%%%%%%%%%%%%%%%
%%%%%%%%%%%%%%%%%%%%%%%%%%%%%%%%%%%%%%%%%%%%%%%%%%%%%%%%%%%%%%%%%

%%%%%%%%%%%%%%%%%%%%%%%%%%%%%%%%%%%%%%%%%%%%%%%%%%%%%%%%%%%%%%%%%
%%%%%%%%%%%%%%%%%%%%%%%%%%%%%%%%%%%%%%%%%%%%%%%%%%%%%%%%%%%%%%%%%

\begin{figure}[hbtp]
\begin{center}
\includegraphics [angle=-90,scale=.75] {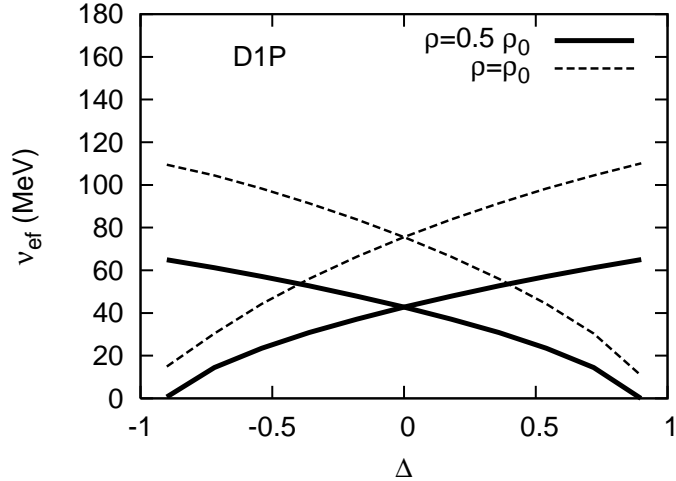}
\caption{Polarized effective neutron chemical potential at densities $\rho=0.5 \rho_0$ (solid line) and $\rho=\rho_0$ (dashed line) at $T=10$ MeV and $B=10^{17}$ G with the Gogny D1P model as a function of the polarization. See text for details.} 
\label{Fig7}
\end{center}
\end{figure}

The composition of the polarized plasma can be obtained by solving the set of 
eqs.(\ref{eq1}), (\ref{eq2}) and (\ref{eq3}). In Fig.~\ref{Fig7} we show the polarized effective neutron chemical potential at densities $\rho=0.5 \rho_0$ (solid line) and $\rho=\rho_0$ (dashed line) at $T=10$ MeV and $B=10^{17}$ G with the Gogny D1P model as a function of the polarization. For each density as polarization grows the ascendent (descendent) lines show the up (down) polarized particle effective chemical potential. As the system is populated with more particles aligned (antialigned) with the magnetic field the energetic cost of aligning a new particle is bigger (smaller). As density increases this cost is also bigger due to nuclear correlations as extracted from the figure.

%%%%%%%%%%%%%%%%%%%%%%%%%%%%%%%%%%%%%%%%%%%%%%%%%%%%%%%%%%%%%%%%%
\begin{figure}[hbtp]
\begin{center}
\includegraphics [angle=-90,scale=.75] {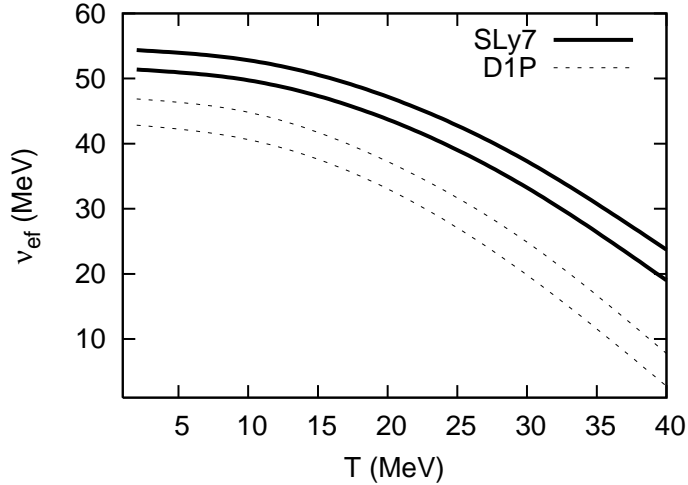}
\caption{Polarized effective chemical potential at a density $\rho=0.5\rho_0$ with the SLy7 model (solid line) and the D1P model (dashed line) and $B=5\,10^{17}$ G  as a function of the temperature.} 
\label{Fig8}
\end{center}
\end{figure}

%%%%%%%%%%%%%%%%%%%%%%%%%%%%%%%%%%%%%%%%%%%%%%%%%%%%%%%%%%%%%%%%%
In order to compare Skyrme and Gogny polarized population yields, in Fig.~\ref{Fig8} we show the polarized effective chemical potential at a density $\rho=0.5\rho_0$ for the SLy7 (solid line) and D1P (dashed line) models with $B=5\,10^{17}$ G as a function of the temperature. Up (down) effective neutron chemical potentials are depicted for each model as the upper (lower) curves. As the system is heated the permanent polarization stays rather constant however the correlations are diminished and the overall energetic cost is smaller.

%%%%%%%%%%%%%%%%%%%%%%%%%%%%%%%%%%%%%%%%%%%%%%%%%%%%%%%%%%%%%%%%%
%%%%%%%%%%%%%%%%%%%%%%%%%%%%%%%%%%%%%%%%%%%%%%%%%%%%%%%%%%%%%%%%%

\begin{figure}
\begin{minipage}[b]{0.5\linewidth} % A minipage that covers half the page
\centering
\includegraphics[angle=-90,scale=0.75]{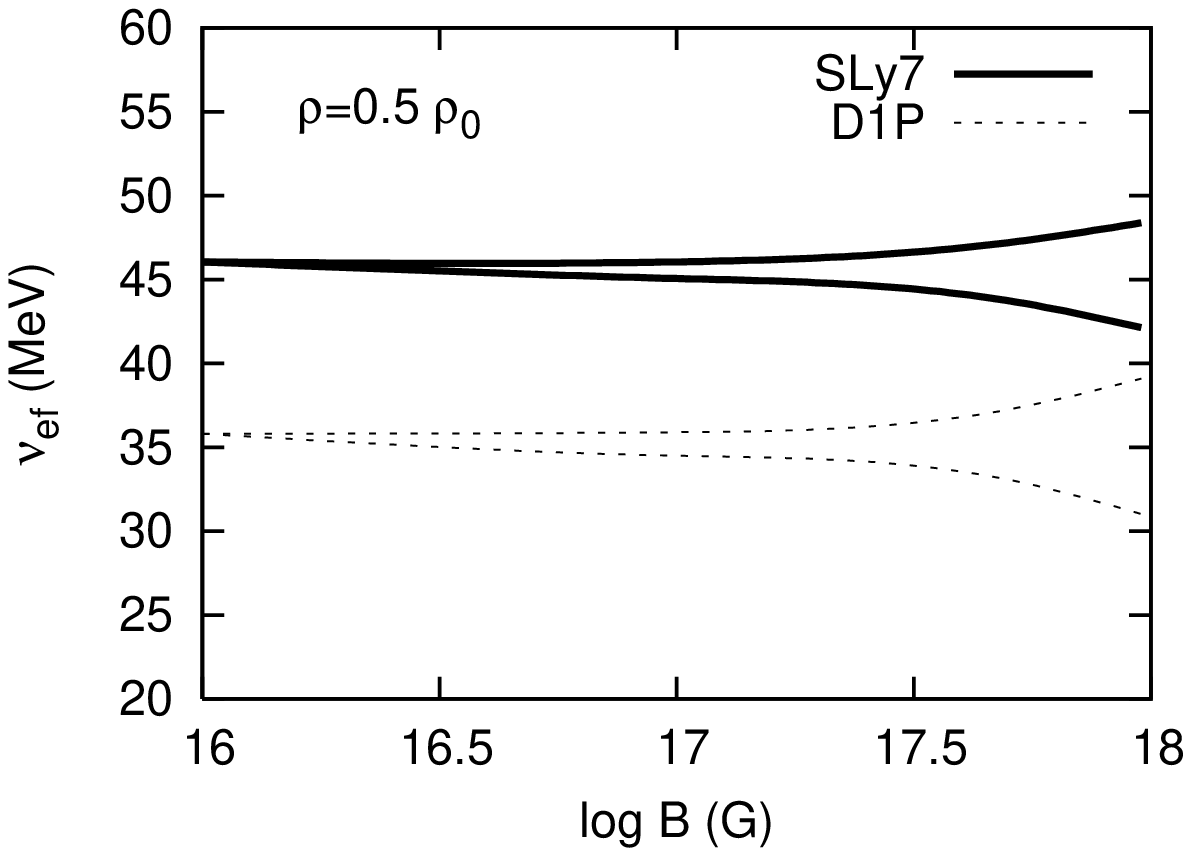}
\end{minipage}
\hspace{0.5cm} % To get a little bit of space between the figures
\begin{minipage}[b]{0.5\linewidth}
\centering
\includegraphics[angle=-90,scale=0.75]{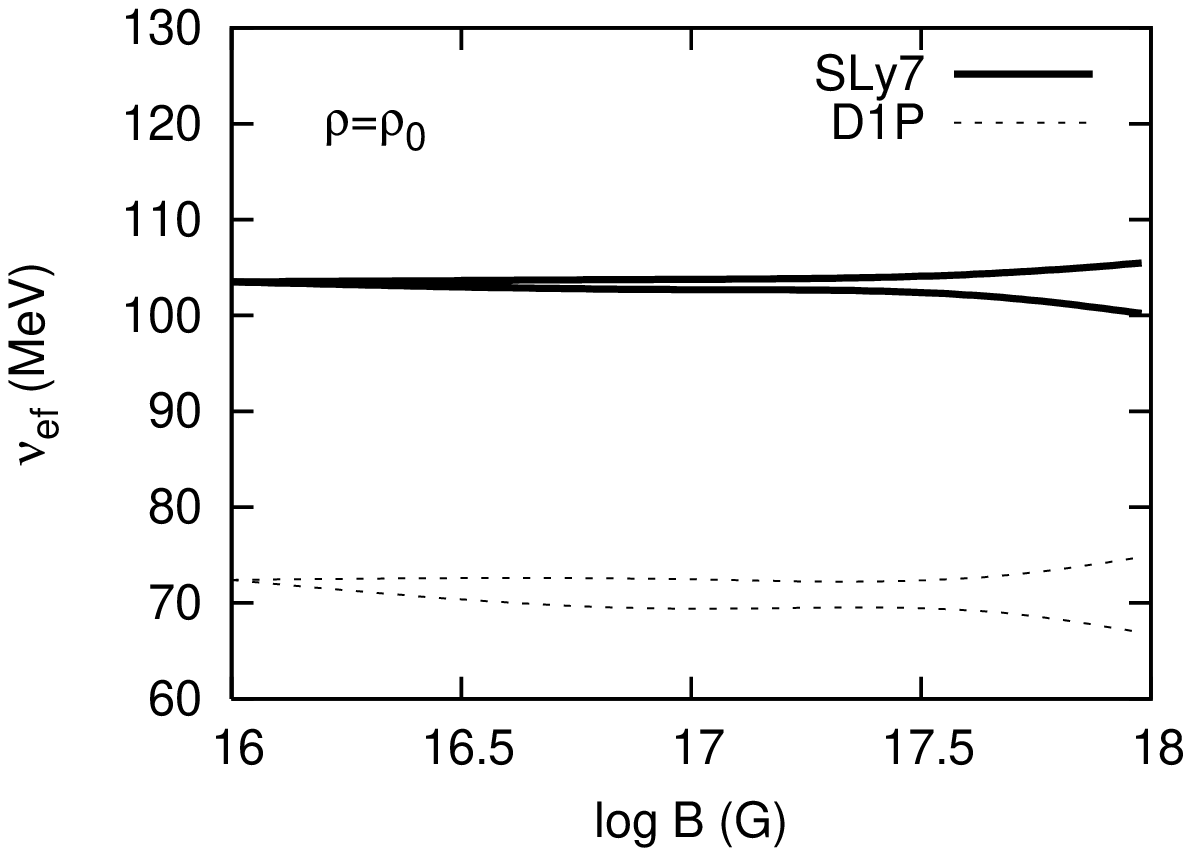}
\caption{Polarized effective chemical potential at densities $\rho=0.5\rho_0$ (upper panel) and $\rho=\rho_0$ (lower panel) with the SLy7 model (solid line) and the D1P model (dashed line) at $T=20$ MeV  as a function of the logarithm of the magnetic field strength. For each model upper (lower) curves correspond to up (down) polarized particles.} 
\label{Fig88}
\end{minipage}
\end{figure}

In Fig.~\ref{Fig88} we show the polarized effective chemical potential densities $\rho=0.5\rho_0$ (upper panel) and $\rho=\rho_0$ (lower panel) with the SLy7 model (solid line) and the D1P model (dashed line) at $T=20$ MeV  as a function of the logarithm of the magnetic field strength. For each calculation model upper (lower) curves correspond to up (down) polarized particles. We can see how the effect of magnetic field strengths smaller than $B\approx 10^{16}$ play almost no role in the plasma polarized neutron population. As $B$ increases there is a gradual enhancement (decrease) of the up (down) polarized neutron sector. However the  parametrization used plays an important role and Skyrme models predict systematically bigger effective neutron chemical potentials than the Gogny forces. From the comparison of upper and lower panels we can see that in-medium effects largely increase the effective chemical potentials.

%%%%%%%%%%%%%%%%%%%%%%%%%%%%%%%%%%%%%%%%%%%%%%%%%%%%%%%%%%%%%%%%%

\section{Summary and conclusions}
\label{summary}

We have studied the effect of a strong magnetic field on some thermodynamical observables in a pure neutron gas within the framework of the non-relativistic Brueckner-Hartree-Fock approximation. We have considered magnetic field strengths up to the maximum value energetically allowed by the scalar virial theorem $B_{max}\approx 10^{18}$ G. We have found that for magnetic field strength values smaller than $B \approx 10^{16}$ G and due to the characteristic tiny value of nucleon magnetic moment the effect on plasma polarization is almost negligible. We have performed a comparative study of the plasma observables using effective Skyrme and Gogny nuclear interaction forces in the density, temperature and magnetic field strength parameter space. With these effective interaction models we find that in the low density limit increasingly large spin polarizations are allowed as density decreases from an upper bound of $\approx 0.5\rho_0$ robustly for both models. Then, for intermediate densities, there is a rather constant polarization plateau where a permanent imanation saturates to a value of up to $12\%$ for the maximum B considered in this work and moderate temperatures. In the high  density region, up to where the meaningful hadronic picture holds, Skyrme and Gogny forces behave very differently. While the Skyrme interaction allows a ferromagnetic transition in the high density limit, the Gogny force prevents it. Temperature effects on magnetization remain moderate in the explored range of temperatures, up to about $T=40$ MeV relevant to proto-neutron stars. The influence of a strong magnetic field should be considered increasingly relevant as density decreases, that is, in the description of the outer shells of proto-neutron stars below $\approx 0.5\rho_0$ where a more detailed analysis must be performed in order to see ther role played by {\it frustration}. For bulk matter in the intermediate and high density range a relatively small polarization is allowed that only affects moderately the thermodynamics. Beta equilibrium should also be appropiately treated since the appearance of electrically charged particles may affect the onset densities of enhanced polarization in the plasma.

%%%%%%%%%%%%%%%%%%%%%%%%%%%%%%%%%%%%%%%%%%%%%%%%%%%%%%%%%%%%%%%%

\vspace{2ex}

\noindent{\bf Acknowledgments}\\
We would like to thank C. Horowitz, J. Navarro and A. Polls for useful discussions.
M.A.P.G. would like to dedicate this work to the memory 
of J.M.L.G. This work has been partially funded by the Spanish Ministry 
of Education and Science project DGI-FIS2006-05319.
%%%%%%%%%%%%%%%%%%%%%%%%%%%%%%%%%%%%%%%%%%%%%%%%%%%%%%%%%%%%%%%%
% Bibliography. 
%%%%%%%%%%%%%%%%%%%%%%%%%%%%%%%%%%%%%%%%%%%%%%%%%%%%%%%%%%%%%%%%

%%%%%%%%%%%%%%%%%%%%%%%%%%%%%%%%%%%%%%%%%%%%%%%%%%%%%%%%%%%%%%%%
\end{document}